\documentclass[12pt]{article}
\usepackage{graphicx}
\usepackage{dcolumn}
\usepackage{bm}
\usepackage{amsmath, amssymb, amsfonts, amsthm}
\usepackage{cite}
\usepackage{epsfig}
\usepackage[usenames,dvipsnames,table]{xcolor}
\usepackage{textcomp}

\def\be{\begin{equation}}
\def\ee{\end{equation}}
\def\ba{\begin{eqnarray}}
\def\ea{\end{eqnarray}}

\def\bdm{\begin{displaymath}}
\def\edm{\end{displaymath}}

\def\bq{\begin{quote}}
\def\eq{\end{quote}}

 at 10truept

\newcommand{\bea}{\begin{eqnarray}}
\newcommand{\eea}{\end{eqnarray}}

\newcommand{\bi}{\begin{itemize}}
\newcommand{\ei}{\end{itemize}}

\newcommand{\beq}{\begin{equation}}
\newcommand{\eeq}{\end{equation}}
\newcommand{\beqa}{\begin{eqnarray}}
\newcommand{\eeqa}{\end{eqnarray}}

\def\ltap{\ \raise.3ex\hbox{$<$\kern-.75em\lower1ex\hbox{$\sim$}}\ }
\def\gtap{\ \raise.3ex\hbox{$>$\kern-.75em\lower1ex\hbox{$\sim$}}\ }
\def\gl{\ \raise.5ex\hbox{$>$}\kern-.8em\lower.5ex\hbox{$<$}\ }
\def\roughly#1{\raise.3ex\hbox{$#1$\kern-.75em\lower1ex\hbox{$\sim$}}}

\begin{document}

\thispagestyle{empty}
\begin{flushright}
May 2017\\
CERN-TH-2017-115\\
DESY-17-080\\

\end{flushright}
\vspace*{.2cm}
\begin{center}
{\Large \bf An \' Etude on Global Vacuum Energy Sequester}\\

\vspace*{.7cm} {\large  Guido D'Amico$^{a, }$\footnote{\tt
damico.guido@gmail.com}, Nemanja Kaloper$^{b, }$\footnote{\tt
kaloper@physics.ucdavis.edu}, Antonio Padilla$^{c, }$\footnote{\tt
antonio.padilla@nottingham.ac.uk},}\\
\vspace{.3cm}
{\large  David Stefanyszyn$^{d, }$\footnote{\tt
d.stefanyszyn@rug.nl}, Alexander Westphal$^{e, }$\footnote{\tt
alexander.westphal@desy.de} and George Zahariade$^{f, }$\footnote{\tt
George.Zahariade@asu.edu}}\\

\vspace{.5cm} {\em $^a$Theoretical Physics Department,
CERN, Geneva, Switzerland}\\
\vspace{.3cm} {\em $^b$Department of Physics, University of
California, Davis, CA 95616, USA}\\
\vspace{.3cm} {\em $^c$School of Physics and Astronomy,
University of Nottingham, Nottingham NG7 2RD, UK}\\
\vspace{.3cm} {\em $^d$Van Swinderen Institute for Particle Physics and Gravity, University of Groningen, Nijenborgh 4, 9747 AG Groningen, The Netherlands}\\
\vspace{.3cm} {\em $^e$Deutsches Elektronen-Synchrotron DESY, Theory Group, D-22603 Hamburg, Germany
}\\
\vspace{.3cm} {\em $^f$Department of Physics, Arizona State University, Tempe, AZ 85287, USA
}\\

\vspace{.5cm} ABSTRACT
\end{center}

Recently two of the authors proposed a mechanism of vacuum energy
sequester as a means of
protecting the observable cosmological constant from quantum radiative corrections. The original proposal
was based on using global Lagrange multipliers, but later a local formulation was
provided. Subsequently other interesting claims of a different non-local approach to the
cosmological constant problem were made, based again on global Lagrange multipliers. We
examine some of these proposals and find their mutual relationship. We explain that the proposals which do not treat the cosmological constant counterterm as a dynamical variable require fine tunings to have acceptable solutions. Furthermore, the counterterm often needs to be retuned at every order in the loop expansion to
cancel the radiative corrections to the
cosmological constant, just like in standard GR. These observations are an important reminder of just how the proposal of vacuum energy sequester avoids such problems.

\vfill \setcounter{page}{0} \setcounter{footnote}{0}
\newpage

\section{Global vacuum energy sequester} \label{prop}

The global vacuum energy sequester of \cite{kalpad1} is based on promoting two gauge-invariant global variables of standard General Relativity (GR) with minimally coupled matter to dynamical degrees of freedom. One of the variables is the cosmological constant counterterm, which in the standard approach is an arbitrary but undetermined variable required to renormalize the cosmological constant. The other is introduced as a dimensionless ratio of the Planck mass and the matter sector mass scales, but can in fact be viewed as the Planck mass counterterm. 
Again, it is an arbitrary but undetermined variable, required to renormalize the Planck scale. Because these quantities are UV sensitive, their numerical values  are determined \textit{ex post facto}, by a measurement, as is usual for any UV sensitive quantity in quantum field theory (QFT) \cite{kalpad1,kalpad2}.

In the ``Einstein conformal gauge", defined by absorbing the Planck scale counterterm into the matter sector via a scale redefinition $\tilde g_{\mu\nu}=\lambda^2 g_{\mu\nu}$, the effective action is
\be
S= \int d^4 x \sqrt{g} \left[ \frac{M^2_{Pl}}{2} R  - \Lambda - {\lambda^4} {\cal L}(\lambda^{-2} g^{\mu\nu} , \Phi) \right] +\sigma\left(\frac{ \Lambda}{\lambda^4 \mu^4}\right)
 \, .
 \label{action}
\ee
Varying (\ref{action}) with respect to $\Lambda, \lambda$ imposes global constraints on the dynamics of the theory.
The action is supplemented with a smooth additive function $\sigma(\Lambda, \lambda)$ which is not integrated over spacetime, such that the variation with respect to $\Lambda$, which yields a constraint for the conjugate variable $\Omega = \int d^4x \sqrt{g}$,  does not force the global world volume $\Omega$ to vanish.

The selection of the global variables $\Lambda, \lambda$ is critical in ensuring that the cosmological constant counterterm automatically cancels the QFT radiative corrections from the source term in the gravitational
field equations.
{For this, it is crucial that the UV regulator of the matter sector introduces contributions where the scales depend on $\lambda$ in precisely the same way as those from the physical fields from ${\cal L}$.}
Furthermore, the general covariance of (\ref{action}) guarantees that,
once chosen, this dependence remains exactly the same at any order of the QFT loop expansion, and so the loop corrections to the cosmological constant will never appear as a source of gravity if the tree level cosmological constant does not gravitate.
Of course, this only accounts for the matter loop corrections from the protected sector.
To cancel gravity loops a slightly more complex proposal involving topological  curvature invariants was presented in \cite{kalpad4}.
For now we shall ignore gravity loops, meaning that gravity is a purely (semi) classical field which merely serves the purpose of {\it detecting} vacuum energy.
This suffices to provide a sharp formulation of the cosmological constant problem \cite{zeldovich,wilczek,ruzi,wein,pad} in gravity as first noted by
Pauli\footnote{However, Pauli refused to lose any sleep over zero point contributions to vacuum energy since he knew he could cancel them by normal ordering \cite{ruzi,akhmedov}. Only with interactions and loops does the problem become a serious one \cite{zeldovich,wein,sirlin}.}.

To see how the loop contributions to the vacuum energy cancel out, we can look at the
field equations that follow from (\ref{action}). The global constraints from varying with respect to $\Lambda$ and $\lambda$ respectively are
\be
\frac{\sigma'}{\lambda^4\mu^4} = \int d^4x \sqrt{g} \, , ~~~~~~~~~~~~~~~~ 4\Lambda \frac{ \sigma' }{\lambda^4\mu^4}
= \int d^4x \sqrt{g} \,  T^{\mu}{}_\mu \, ,
\label{varsl}
\ee
where $ T_{\mu\nu}=-\frac{2}{\sqrt{ g}} \frac{\delta S_m}{\delta g^{\mu\nu}}$ is the canonical energy-momentum tensor. Here $\sigma' = \frac{d\sigma(z)}{dz}$ and as long as $\sigma'$ is non-zero and non-degenerate \cite{kalpad1,kalpad2},  the constraints are invertible and yield
 \be \label{ccfix}
 \Lambda = \frac14 \langle T^\mu{}_\mu \rangle \, .
 \ee
We have defined the  $4$-volume average of a quantity by $\langle Q\rangle ={\int d^4 x \sqrt{g} \, Q}/{\int d^4 x \sqrt{g}}$, noting that such averages are delicate and must be defined carefully \cite{kalpad1,kalpad2}.
Equation (\ref{ccfix}) completely fixes the cosmological constant counterterm $\Lambda$ in terms of the  matter sources. This is because we treat $\Lambda$  as a dynamical variable whose value is determined by equation (\ref{ccfix}).

Substituting $\Lambda$ into the gravitational field equations yields
\be
M_{Pl}^2 G^\mu{}_\nu = -\Lambda \delta^\mu{}_\nu +  T^\mu{}_\nu = T^\mu{}_\nu-\frac{1}{4} \delta^\mu{}_\nu \langle   T^\alpha{}_\alpha \rangle \, .
\label{eeqs}
\ee
Here, unlike in unimodular gravity~\cite{finkel, buch, henteitel, unruh, ng, kuchar, griff}, there are \emph{no} hidden equations nor integration constants,  and all the sources are automatically accounted for in~\eqref{eeqs}.
The counterterm $\Lambda$ is a global dynamical field  fixed by the field equations. Crucially, $ \langle  T^\alpha{}_\alpha \rangle / 4$
is subtracted from the right-hand side of (\ref{eeqs}) meaning that the hard cosmological constant, be it a classical contribution to ${\cal L}$ in (\ref{action}) or a quantum vacuum correction calculated to any order in the loop expansion, divergent (but regulated!) or finite, {\it never} contributes to the field equations (\ref{eeqs}).

Indeed, we can take the matter Lagrangian ${\cal L}_\text{eff}$ at any given order in loops, split it into the renormalized quantum vacuum energy contributions $\tilde V_{vac}=\langle 0| {\cal L}_\text{eff} (\tilde g^{\mu\nu}, \Phi )|0\rangle$, and local excitations $\Delta {\cal L}_\text{eff}$,
\be
\lambda^4 \sqrt{g} {\cal L}_\text{eff}(\lambda^{-2} g^{\mu\nu} , \Phi)=\lambda^4 \sqrt{g}\left[ \tilde  V_{vac} + \Delta {\cal L}_\text{eff}(\lambda^{-2} g^{\mu\nu} , \Phi)\right]
\ee
to find that $T^{\mu}{}_\nu =- V_{vac} \delta^\mu{}_\nu + \tau^\mu{}_\nu$, where $V_{vac}=\lambda^4\tilde V_{vac} $ is the total regularized vacuum energy and $\tau_{\mu\nu}= \frac{2}{\sqrt{g}} \frac{\delta}{\delta g^{\mu\nu}} \int d^4x \sqrt{g} \lambda^4 \Delta {\cal L}_\text{eff}(\lambda^{-2} g^{\mu\nu}, \Phi)$ describes local excitations.
Since the average of a constant is just the constant itself, $\langle V_{vac} \rangle \equiv  V_{vac}$, the field equations (\ref{eeqs}) become
\be
M_{Pl}^2 G^\mu{}_\nu=  \tau^\mu{}_\nu-\frac{1}{4} \delta^\mu{}_\nu \langle  \tau^\alpha{}_\alpha \rangle \, .
\label{eeqs1}
\ee
The regularized vacuum energy $V_{vac}$  has completely dropped out from the source in (\ref{eeqs}). The residual effective cosmological constant arises from the historic average of the trace of matter excitations:
\be
\Lambda_\text{eff} =\frac14 \langle \tau^\mu{}_\mu \rangle \, .
\label{residual}
\ee
It turns out that $\Lambda_\text{eff}$ is automatically small enough in large old universes, as shown in \cite{kalpad1,kalpad2}.
Importantly, it has absolutely nothing to do with the vacuum energy contributions from the matter sector, including the Standard Model contributions.
Instead, after the vacuum energy drops out by virtue of the global constraints, the residual value of $\langle \tau^\mu{}_\mu \rangle$ is picked by the total cosmological evolution. This measurement  requires the whole history of the universe to determine this variable precisely \cite{degrav,kalpad1,kalpad2}.

A subtlety of this formulation of global sequester is that we must have $\lambda \ne 0$, since $\lambda \propto m_{phys}/M_{Pl}$.
The solutions where $\lambda = 0$ force the QFT to the conformal limit, where the theory effectively has no dimensionful parameters.
The first of equations (\ref{varsl}) shows that if $\lambda$ is non-zero, $\int d^4x \sqrt{g}$ must be {\it finite}.
This picks out a universe with spatially compact sections, which is also temporally finite: it starts with a Bang and ends with a Crunch.
Infinite universes are solutions too, however their phenomenology would not be a good approximation to our world, since all scales in the protected sector vanish.
Local formulations of vacuum energy sequester \cite{kalpad3,kalpad4} evade these restrictions.

It is instructive to look at the global sequester theory in the ``Jordan'' conformal gauge. Defining
$
\tilde g_{\mu\nu} =\lambda^2 g_{\mu\nu}, \tilde \Lambda=\frac{\Lambda}{\lambda^4}$ and $\kappa^2=\frac{M_{Pl}^2}{\lambda^2}
$, the action reads
\be
S= \int d^4 x \sqrt{\tilde g} \left[ \frac{\kappa^2}{2} \tilde R  - \tilde \Lambda - {\cal L}(\tilde  g^{\mu\nu} , \Phi) \right] +\sigma\left(\frac{ \tilde \Lambda}{ \mu^4}\right)
 \, .
 \label{jaction}
\ee
This manifestly shows that the UV sensitive couplings of GR -- the cosmological constant and the Planck scale -- are promoted to global dynamical variables ($\tilde \Lambda$ and $\kappa$ respectively).
Note the specific role played by $\kappa$ which is very clear in this gauge.
Its variation fixes the spacetime average of the Ricci scalar to vanish.
Along with Einstein's equations and the cosmological constant counterterm variable, this global geometrical constraint guarantees the dynamical cancellation of vacuum energy loops. Our choice is not unique -- all we need is a global constraint that ties a scale dependent curvature invariant to an IR observable.
For example, in the local formulation of vacuum energy sequester the spacetime average of the Ricci scalar is tied to the fluxes of 4-forms which are arbitrary IR quantities, that can be naturally small \cite{kalpad3}.
A generalization that removes the graviton loop contributions to vacuum energy constrains the spacetime average of the Gauss-Bonnet invariant, by again fixing it to the flux of a 4-form \cite{kalpad4}.

An added bonus of this gauge is that it is manifest to see that the function $\sigma$ must be non-linear.
A linear $\sigma$ would yield two geometrical constraints, $\int d^4 x \sqrt{\tilde g}\tilde R=0$ and $\int d^4 x \sqrt{\tilde g}=1/\mu^4$.
The first of these can be satisfied, via Einstein's equation, by dynamically fixing the value of $\tilde \Lambda$.
However, after $\kappa^2$ is fixed by matching Einstein's equation to local gravitational experiments, to satisfy the second constraint one must fine tune other integration constants.
For example, to accommodate a large and old universe we would have to take $\mu$ to lie at scales many orders of magnitude below particle physics scales and the cut-off of our effective field theory.
For non-linear choices of $\sigma$, such tunings are avoided.

\section{Volumes, $\hbar$'s, stiff dilatons and all that}

Other formulations that utilize global constraints in gravity exist.
A standout among them is the attempt by  Tseytlin
to formulate a low energy effective theory of gravity based on ideas about a manifest T-duality invariant formulation of target space actions in string theory,
\cite{tsey}, inspired by ideas of \cite{linde2}. While such formulations are under development, Tseytlin noticed that in such classes of theories, if they exist, and if they are applied to very asymmetric compactifications of the
winding mode variables, the leading order action in the IR can be written as
\be
S_T = \frac{{\int d^4x \sqrt{g} \left[\frac{M_{Pl}^2}{2} R - {\cal L}_0 -  {\cal L}(g^{\mu\nu}, \Phi) \right]}}{\left[{\mu^4 \int d^4x \sqrt{g}}\right]}.
\label{tseytact}
\ee
Clearly, the classical and zero-point (tree-level) contributions to the cosmological constant ${\cal L}_0$ immediately drop out from the gravitational field equations obtained by varying this action, since they are $\propto \int d^4x \sqrt{g}$, and this term is cancelled by the denominator. So whatever they are, they simply do not gravitate\footnote{This is really not much of an accomplishment since the same feat can be achieved with normal ordering. Again, the problem is with loops.}. However, this does
not eliminate the loop contributions to the vacuum energy. A very useful and simple way to see this is to note that
if one formulates a quantum theory based on the action (\ref{tseytact}), the worldvolume in the denominator acts
just like $\hbar$. In QFT, the powers of $\hbar$ count the loop corrections to the effective action, and the loop expansion is organized as
\be
 S_{eff}  = \frac{S_0}{\hbar} + S_1 + \hbar S_2 + \hbar^2 S_3 + \ldots \, ,
\ee
and so on. Thus if $S_T$ in (\ref{tseytact}) is to be used as the starting point for formulating a QFT coupled to
gravity, the full effective action associated with it would be of the form
\be
S^T_{eff} = \frac{S^T_{0}}{\Omega} + S^T_{1} + \Omega S^T_{2} + \Omega^2 S^T_3  + \ldots \, ,
\label{tseyeff}
\ee
where $\Omega = \mu^4 \int d^4x \sqrt{g}$. Obviously, the cosmological constant contributions from the corrections are large
in any conventional local QFT, generically being of the order of the $({\rm cutoff})^4$ unless there is a dynamical principle like supersymmetry or conformal symmetry to suppress them. In other words, because the theory has UV sensitive quantities, the ``corrections" aren't small despite the fact that they are ``higher order". This is the essence of the radiative stability problem. Further, their dependence on the
worldvolume $\Omega = \mu^4 \int d^4x \sqrt{g}$ is {\it different} than the classical and tree level term, and
they would {\it not} cancel from the gravitational field equations obtained by varying (\ref{tseyeff}).

A more precise way to see this, and also to pursue the contact with vacuum energy sequester, is to
rewrite the theory (\ref{tseytact}) as
\be
S _{T} = \int d^4x \sqrt{g} \left[\frac{\hat \lambda^4 M_{Pl}^2}{2} R -\hat  \Lambda - \hat \lambda^4 {\cal L}(g^{\mu\nu}, \Phi) \right] + \frac{\hat \Lambda}{\hat \lambda^4 \mu^4} \, ,
\label{tsey}
\ee
with the introduction of the global Lagrange multiplier variables $\hat \Lambda, \hat \lambda$ (where we have absorbed
${\cal L}_0$ into ${\cal L}$).
One can readily verify that integrating out $\hat \Lambda, \hat \lambda$ yields precisely Tseytlin's action (\ref{tseytact}).
This is indeed reminiscent of the global vacuum energy sequester (\ref{action}). However the key difference is the dependence of the bulk terms on $\hat \lambda$.
Here, the Einstein-Hilbert term has a $\hat \lambda^4$ prefactor, and the matter sector does not have the kinetic energy scaling $\propto 1/\hat \lambda^2$.
Normalizing the Einstein-Hilbert term canonically, $g_{\mu\nu} \to \hat \lambda^{-4} g_{\mu\nu}$, and defining the new variables $\lambda = 1/\hat \lambda, \Lambda=\hat \Lambda/\hat \lambda^8$, yields
\be
S _T = \int d^4x \sqrt{g} \left[\frac{M_{Pl}^2}{2} R - \Lambda - { \lambda^4} {\cal L}\left(\lambda^{-4}g^{\mu\nu}, \Phi\right) \right] + \frac{\Lambda }{\lambda^4 \mu^4}
\, .
\label{tseactconst}
\ee
The tree-level vacuum energy $V_{vac} = {\cal L}_0 \lambda^4$ scales like $\lambda^4$ and indeed it will be automatically eliminated from the source of Einstein's equations once $\lambda $ is integrated out, as can be seen from the field equations obtained by varying~\eqref{tseactconst}~\cite{kalpad1,kalpad2}.
The $\lambda$-dependence of the matter sector and the counterterm $\Lambda$ is engineered precisely to accomplish this.
That is evident by integrating $\lambda,\Lambda$ out, after which \eqref{tseactconst} reverts back to \eqref{tseytact}.

However, after canonically normalizing the QFT Lagrangian, the physical masses scale as $m_{phys}=m \lambda^2$, and so the radiative corrections to the vacuum energy scale as $\lambda^8$. Thus they will not automatically cancel, and will restore the vacuum energy radiative instability
exactly as in GR or unimodular gravity \cite{kalpad1,kalpad2}. This $\lambda$ dependence does not correctly count the engineering dimension of the vacuum energy loop
corrections, unlike in the vacuum energy sequester proposal.
Further, as already commented in the {\it Note added} in \cite{tsey}, the Planck mass is also radiatively unstable, receiving corrections
$\Delta M_{Pl}^2 \simeq {\cal O}(1) \times m_{phys}^2 \simeq {\cal O}(1) m^2  \lambda^4$, which are large in old and large universes where $\hat \lambda = 1/\lambda$ is small.
The dynamics of vacuum energy sequester based on (\ref{action}) is designed to get around this problem, by fixing the scalings with $\lambda$ in the matter Lagrangian, as we explained in the previous section. We also promote the global term $ \frac{\Lambda }{\lambda^4 \mu^4} $ to a more general non-linear function $\sigma \left( \frac{\Lambda }{\lambda^4 \mu^4} \right)$ to avoid an implicit fine tuning. We note that the higher-dimensional  multi-tensor framework of \cite{gabadadze} motivated by Tseytlin's theory has a similar problem in that the radiative corrections do not obey the same form of the Lagrange multiplier dependence as the original action for generic values of the global variables. 
However, this proposal can also be  modified by  bringing it into the form of a generalized model of vacuum energy sequestering as shown in section \ref{sec:local}.

An alternative approach has been pursued in \cite{westphal}, where the authors in effect take the global sequester action (\ref{action}), drop the global term $\sigma$, replace the cosmological constant counterterm of global sequester $\Lambda$ by another
completely arbitrary constant ${\cal L}_0$, which unlike $\Lambda$ scales with a stiff dilaton $\lambda$ as $\lambda^4$ and treat it as the scaled cosmological constant counterterm.
The action is
\be
S_{BRRW} = \int d^4 x \sqrt{g} \left[ \frac{M^2_{Pl}}{2} R  - \lambda^4 {\cal L}_0 - {\lambda^4} {\cal L}(\lambda^{-2} g^{\mu\nu} , \Phi) \right]
 \, .
 \label{actionwest}
\ee
Again, ${\cal L}_0$ is a completely arbitrary quantity, but \cite{westphal} {\it do not} vary the action with respect to it.
This theory looks very simple in the ``Jordan" gauge. Using
$\tilde g_{\mu\nu} = \lambda^2 g_{\mu\nu}$,
\be
S _{BRRW} = \int d^4x \sqrt{\tilde g} \left[\frac{M_{Pl}^2}{2 \lambda^2} \tilde R - {\cal L}_0 - {\cal L}(\tilde g^{\mu\nu}, \Phi) \right] \, .
\label{west}
\ee
The variations yield $\lambda^{-2} M_{Pl}^2 \tilde G^{\mu}{}_\nu = \tilde T^{\mu}{}_\nu - {\cal L}_0 \, \delta^{\mu}{}_{\nu}$ and  $\langle \tilde R \rangle  = 0$.
Taking the trace of Einstein's field equations, averaging over spacetime and using
$\langle \tilde R \rangle = 0$ yields ${\cal L}_0 = \langle \tilde T^\mu{}_\mu \rangle/4$. Since ${\cal L}_0$ is not varied over, ${\cal L}_0 = \langle \tilde T^\mu{}_\mu \rangle/4$ is a \emph{consistency condition} that needs to be satisfied \emph{by choosing an appropriate boundary condition} so that tracing and averaging Einstein's equations reproduces the $\lambda$ equation of motion $\langle\tilde R\rangle=0$. Thus
the field equations are equivalent to
\be
{\cal L}_0 = \frac14 \langle \tilde T^\mu{}_\mu \rangle \, , ~~~~~~ \lambda^{-2} M_{Pl}^2 \tilde G^{\mu}{}_\nu = \tilde T^{\mu}{}_{\nu} - \frac14  \langle \tilde T^\lambda{}_\lambda \rangle \, \delta^{\mu}{}_{\nu} \, .
\ee
These equations are formally identical to the sequester field equations (\ref{ccfix}) and (\ref{eeqs}). Thus it is very tempting to interpret the theory (\ref{west}) as a form of vacuum energy sequester with an additive action. This could
lend to a simpler means of quantization and perhaps a more straightforward road to UV embeddings.

There is a serious obstruction to this interpretation. The quantity ${\cal L}_0$, which plays the role of the cosmological constant counterterm here, is {\it not} a dynamical variable in this theory, but an integration constant
\footnote{By an {\it integration constant} we mean any arbitrary parameter whose value is fixed by boundary conditions.
With this terminology, one should even refer to the cosmological counterterm of GR as an integration constant since it is a priori arbitrary but fixed by its value specified in the boundary data.
In unimodular gravity \cite{finkel,buch,henteitel,unruh, griff}, the cosmological constant counterterm is often bizarrely celebrated for being an integration constant, although it should be obvious that this is identical to the situation in GR.}
whose value is determined by arbitrary boundary conditions. Since as in any theory such boundary conditions are supplied by observation, this means that the counterterm needs to be matched to reproduce the measurement of whatever the vacuum curvature of the universe is.
Thus satisfying ${\cal L}_0 = \langle \tilde T^\mu{}_\mu \rangle /4$ means that one must pick the value of this integration constant \emph{by hand}, or fine tune it precisely to the value of $ \langle \tilde T^\mu{}_\mu \rangle /4$, and redo this fine tuning -- again, by hand -- order by order in perturbation theory.
Note, that the $\lambda$-scaling of the matter sector in \eqref{west} is the same as in full sequestering. Therefore, the matter loop contributions to the vacuum energy preserve the $\lambda$-dependence found for full sequestering: they are form invariant under $\lambda$-scaling, yet additive and of comparable magnitude at each loop order, set by the scale ${\cal M}_{UV}$ of breaking of full scale symmetry in the UV.
This is similar to the finite corrections to the electro-weak hierarchy arising in softly broken supersymmetry at all loop orders.

Since there is a large hierarchy between the vacuum curvature $\simeq 1/H$ and the scale of scaling symmetry breaking ${\cal M}_{UV}$, the theory \eqref{west} is not radiatively stable because the cosmological constant counterterm $\Lambda$ can drift arbitrarily far as loop corrections are summed over. This is unlike  the dynamics of full sequester where the counterterm is a derived quantity set by the worldvolume of the universe, or the topological fluxes through it,
which are IR quantities. Further problems related to the difficulties of protecting $\lambda$ from developing its own local dynamics, as well as the issues with emerging conformal symmetry and its breaking were also noted in~\cite{westphal}.

In vacuum energy sequester \cite{kalpad1,kalpad2} the dynamics picks the correct value of the vacuum energy {\it variable} $\Lambda$ automatically at every order in perturbation theory.  This places no additional tunings on the matter sector, since  $\Lambda$ is allowed to vary globally.  One can think of $\Lambda$ as satisfying Neumann boundary conditions, which automatically holds by virtue of the field equation $\partial \Lambda = 0$ following from identifying $\Lambda$ with a dual of the flux of a 4-form field strength \cite{henteitel,kalpad3}. More precisely, this field equation arises as a consequence of the gauge symmetry of the 4-form which imposes that $\Lambda$ is a global degree of freedom. The fact that the boundary condition can be satisfied trivially is not really surprising
since $\Lambda$ appears as the conjugate momentum to the purely spatial part of a three-form so the action is a Routhian. It is manifestly in first-order form with respect to this canonical pair and thus its variation doesn't need to be constrained at the boundary.

Conversely if the proposal of \cite{westphal} is altered to allow for global variations in ${\cal L}_0$, one must
include an additional global term in the action to avoid forcing spacetime volume to vanish. The simplest example of this is a non-linear global term of the form $\sigma\left(\frac{{\cal L}_0}{ \mu^4 }\right)$, which would result in the theory being identical to global vacuum energy sequester.

Very recently another apparently different proposal was given in \cite{carrollremmen}. Briefly, these authors consider a theory
defined by the action
\be
S _{CR} = \eta \int d^4x \sqrt{\tilde g} \left[\frac{M_{Pl}^2}{2} \tilde R - {\cal L}_0 - {\cal L}(\tilde g^{\mu\nu}, \Phi) -
\frac{1}{48}  \tilde F_{\mu\nu\lambda\sigma}^2 + \frac16  \tilde \nabla_\mu(\tilde F^{\mu\nu\lambda\sigma} \tilde A_{\nu\lambda\sigma}) \right] \, .
\label{cr}
\ee
The motivation for this theory is that this is essentially GR with an alternate measure, given by
some arbitrary 4-form $H = \eta \sqrt{\tilde g} d^4x$, where this identity simply follows from the fact that any
4D 4-form is proportional to the standard volume element. Thus introducing such a structure brings in one single degree of freedom, the magnetic dual $\eta$. This quantity is postulated to be a global variable for which the action should be varied over. Note that this 4-form is {\it not} a gauge-field strength but just a stationary `potential' without any local dynamics.

The role of the other 4-form $\tilde F$ is subtle. If one imagines that this 4-form arises in response to membrane sources
in the theory, which is a common case with the $p$-forms in string theory, then this form is just an additional `normal' degree of freedom included in the matter sector and can be treated in precisely the same way as any other local matter field. Its flux is uniquely determined by the
distribution of membrane sources in spacetime, and its magnetic dual flux $\tilde F = \theta \sqrt{\tilde{g}} d^4x$, which is a constant by virtue of the equations of motion $d\theta = 0$, just shifts the vacuum energy density, ${\cal L}_0 \rightarrow {\cal L}_0 + \frac{\theta^2}{2}$.
The quantity ${\cal L}_0$ again plays the role of the cosmological constant counterterm, and just as in the approach of \cite{westphal} is an arbitrary integration constant but not varied over in the action (or equivalently integrated over in the path integral).
Canceling the cosmological constant by picking its value again
represents fine tuning.

Imagine however that the 4-form, $\tilde F$, has no membrane sources in the theory.
Let it be a field strength associated with a 3-form  potential $\tilde A$, which arises as a topological property of the manifold. Again, $\tilde F$ can be dualized and its flux shifts the vacuum energy density by  ${\cal L}_0 \rightarrow {\cal L}_0 + \frac{\theta^2}{2}$.
However, in the absence of sources for it, this term is a global degree of freedom, just as in so-called q-theory models~\cite{qth}.
Its value is completely arbitrary and so it can be freely used to solve the constraint equations of the theory, and adjust vacuum energy away. We will focus on this reinterpretation of~\eqref{cr} since the case with sources is obviously fine tuned in the absence of any other dynamics.
In fact we will see that this reinterpretation really links this theory to the local formulation of vacuum energy sequester \cite{kalpad3,kalpad4}.

To explain these points, let's consider the dualities which relate the actions with variables $\tilde F$ and its magnetic dual $\theta$. We can use the Lagrange multiplier method to go between conjugate variables (the methodology is identical to dual reformulations of flux monodromy models \cite{flux}).
Ignoring the surface terms without any loss of generality, we can start with the magnetic dual replacing the matter Lagrangian by ${\cal L}_0 \rightarrow {\cal L}_0 + \frac{\theta^2}{2}$, and adding to the total action a Lagrange multiplier of the form, $-\eta\mu^2 \sigma \left(\theta  \int d^4 x \sqrt{\tilde g}-\int \tilde F\right)$. Integrating out $\sigma$ leaves the magnetic dual action in terms of $\theta$. However, if we first
integrate out $\theta$, which yields $\theta=-\mu^2 \sigma$, we find
\be
S _{CR} = \eta \int d^4x \sqrt{\tilde g} \left[\frac{M_{Pl}^2}{2} \tilde R - {\cal L}_0 +\frac12 \mu^4 \sigma^2 - {\cal L}(\tilde g^{\mu\nu}, \Phi) \right]+\eta\mu^2 \sigma \int \tilde F \, .
\label{crseq}
\ee
If we next integrate out $\sigma$, the resulting action is precisely the one used by \cite{carrollremmen}, with the standard 4-form kinetic term $- \tilde F^2/48$.
On the other hand the ``hybrid" action (\ref{crseq}) is very useful since it established a link with the local vacuum energy sequester.

Indeed, if we rescale the 4-form to absorb the factor of $\mu^2$, $\tilde F \to \tilde F/\mu^2$, and field-redefine the global variable\footnote{This field redefinition is non-analytic at $\tilde \Lambda = 0$ in terms of original variables since changing the sign of $\Lambda$ requires a complex rotation of $\sigma$.}  $\tilde \Lambda=-\frac12 \mu^4 \sigma^2$, we obtain
\be
S _{CR} = \eta \int d^4x \sqrt{\tilde g} \left[\frac{M_{Pl}^2}{2} \tilde R -\tilde \Lambda - {\cal L}(\tilde g^{\mu\nu}, \Phi) \right]
+ \eta \sigma\left(\frac{\tilde \Lambda}{\mu^4}\right) \int \tilde F \, ,
\label{crseq1}
\ee
where $\sigma^2(z)=-2z$.
We have also absorbed ${\cal L}_0$ into ${\cal L}$. Note, that while we have worked with a specific form of the
$\sigma$ function, dictated by the initial quadratic dependence of the action of $\tilde F$, we could have used an arbitrary function instead. This action is reminiscent of a hybrid of global and local vacuum energy sequester.
To see the connection to the local sequester, note that $\tilde \Lambda$ can be viewed as a local degree of freedom (inside the integral), whose local fluctuations however are pure gauge by virtue of the $\tilde F$ term.
The variable $\eta$ remains global\footnote{This variable can be promoted to a local variable following the approach of the local vacuum energy sequester \cite{kalpad3,kalpad4}, as shown recently by \cite{oda}. However
that does not help with the problem of radiative stability.}.

However, the main problem of \cite{carrollremmen} is that the vacuum energy cancellation is not radiatively stable.
While adjusting sourceless $\theta$ can cancel a classical/tree-level cosmological constant, the same procedure will fail to cancel quantum corrections in the loop expansion.
Indeed, just like in the case of Tseytlin's action, the global Lagrange multiplier is an overall factor in the action, essentially corresponding to  $1/\hbar$.
{As a result, if the action \eqref{cr} is to be understood as a zeroth order term in the perturbative expansion of the full quantum effective action, such that we can use it as the phase of a path integral weight, the corresponding theory is not radiatively stable}.
This follows since once the tuning of the cosmological constant is done at tree level, it needs to be redone systematically and very severely to continue canceling the higher loop corrections.
As we discussed above, higher loop terms come with different powers of $\eta = 1/\hbar$, and so the field equations change dramatically when these are included.

To see this explicitly we pass to the `Einstein' gauge by rescaling the metric  $\tilde g_{\mu\nu} = g_{\mu\nu}/\eta$  and the 4-form field strength $\tilde F = F/\eta$, and define $\lambda = \eta^{-1/4}$, $\Lambda=\tilde \Lambda/\eta$, so that
 \be
S _{CR} =  \int d^4x \sqrt{ g} \left[\frac{M_{Pl}^2}{2} R - \Lambda - \lambda^4 {\cal L}(  \lambda^{-4}  g^{\mu\nu}, \Phi)  \right]+\sigma\left(\frac{ \Lambda}{\lambda^4 \mu^4}\right) \int  F  \, .
\label{cr1}
\ee
By comparing with (\ref{tseactconst}), we see that the matter Lagrangian shares the same $\lambda$ dependence as the reduction of Tseytlin's action.  From this $\lambda$ dependence and the arguments regarding radiative stability of Tseytlin's action it is clear that \eqref{cr} is not radiatively stable if treated as the starting point for QFT perturbation theory even with the reinterpretation of the $4$-form flux as a free variable, since the loop corrections will depend differently on $\lambda$ than the tree-level action. So as with Tseytlin's model, the proposal of \cite{carrollremmen} fails to cancel the loop contributions to vacuum energy.  
Thus this theory really does not help with the cosmological constant problem.

Nevertheless this problem can be rectified, by using the lesson from the mechanism of vacuum energy sequester.
In the proposal \cite{carrollremmen} the global constraint arising from the variation with respect to $\eta$ is not purely geometrical. It includes a contribution from the matter action. As an illustration, consider the constraint in the absence of local matter excitations, so that we only have vacuum energy as a source.
This now takes the form of $\frac{M_{Pl}^2}{2} \langle \tilde R \rangle-\tilde \Lambda_\text{eff}=0$, where $\tilde \Lambda_\text{eff}$ is the renormalized cosmological constant containing the tree level vacuum energy and the globally varying cosmological constant counterterm. It follows from $M_{Pl}^2 \tilde G_{\mu\nu}=-\tilde \Lambda_\text{eff} \tilde g_{\mu\nu}$ that $\tilde \Lambda_\text{eff}=0$.
However this constraint is {\it not} robust against loop corrections which spoil the desired cancellation.
To cure this, one needs to alter the theory such that the global constraint is purely geometrical, such as $\langle \tilde  R \rangle =0$. A simple example is
\be
S_{hybrid} = \int d^4x \sqrt{\tilde g} \left[\frac{M_{Pl}^2}{2}\eta  \tilde R - {\cal L}_0 - {\cal L}(\tilde g^{\mu\nu}, \Phi) -
\frac{1}{48}  \tilde F_{\mu\nu\lambda\sigma}^2 + \frac16  \tilde \nabla_\mu(\tilde F^{\mu\nu\lambda\sigma} \tilde A_{\nu\lambda\sigma}) \right] \, ,
\label{crmod}
\ee
where the Lagrange multiplier only multiples the Ricci scalar part and not the entire action. By performing the 4-form manipulations as in \eqref{crseq} and \eqref{crseq1}, we see that this is really hybrid vacuum energy sequester, where the local terms for the Lagrange multiplier $\eta = \eta(\lambda)$ which are proportional to the 4-form \cite{kalpad3} that dynamically enforces $d \eta = 0$ are manifestly dropped.
In this theory, $\lambda$ is treated as a global variable, while $\Lambda$ is local, but with all of its fluctuations
projected out by the gauge symmetry of its dual $4$-form. The loop corrections are automatically cancelled at any order in the loop expansion,  and simultaneously the renormalized vacuum energy is picked to be zero, by virtue of $\langle \tilde R \rangle = 0$.
We will discuss such hybrids in more detail below.

\section{Nonperturbative speculations}

The discussion of the previous section clearly demonstrates the problems which one encounters when introducing global constraints to address the cosmological constant problem.
To avoid direct fine tuning of the renormalized cosmological constant, the cosmological constant counterterm must be treated as a dynamical variable, which in principle has its conjugate momentum and its own equation of motion. Appropriate dynamics may be constructed to select the correct value of the counterterm as an on-shell condition as opposed to having to choose its value by hand.
This could be avoided by invoking the anthropic principle \cite{wein}, but then one is  back to the landscape arguments.
Another issue is that the Lagrange multipliers must be introduced carefully to ensure that the on-shell conditions -- the field equations which enforce the correct value of the cosmological constant counterterm -- are radiatively stable.
Otherwise, anything gained at the classical level is lost once quantum corrections are included.
Further, to ensure that the global constraints are meaningful, one may need to add global contributions to the action which seem to obstruct their use in a path integral.
For example, the global term in the action $\sigma\left(\frac{\Lambda}{\lambda^4 \mu^4}\right)$ prevents the resulting constraint from enforcing the vanishing of the 4-volume, $\Omega = 0$ \cite{kalpad1,kalpad2}.

Since at some point one has to consider quantizing theories which sequester, one needs to have a better understanding of such non-additive actions.
They might arise in attempts to model non-perturbative dynamics of gravity, for example using wormholes \cite{coleman}.
However since the wormhole calculus is fraught with its own difficulties \cite{wormholes1,wormholes2} it is hard to come up with explicit examples which are completely under control. Even in Tseytlin's approach \cite{tsey}, the action which can be interpreted as a theory with a global constraint is proposed to arise by integrating out UV degrees
of freedom and imposing $T$-duality on the result. However in that case there does not yet appear to be a scheme which would reproduce the correct radiative stability of the ensuing low energy action. Such a scheme, if it exists, should yield a different dependence of the higher loop corrections on the compactification parameters than the
leading order action, to restore radiative stability. An elegant resolution of difficulties with non-additivity of the action is to replace the global constraints by local ones, using enhanced gauge symmetries of additional topological 4-form sectors, as introduced in \cite{kalpad3,kalpad4} following the covariant formulation of unimodular gravity \cite{henteitel}.
We will review this procedure in the next section.

One might try to evade this problem by promoting the actions into the full effective actions with all the loops included \cite{ng,smolin}, so that the problem simply disappears \cite{gabadadze}.
However, one then needs to explain how such theories arise from some standard QFT, such as the Standard Model.
In other words, one needs a quantized version of the theory where the corrections to vacuum energy that are higher power in $\hbar$ (and hence higher order in the loop expansion) are automatically parametrically small.
Except for vacuum energy sequester, as far as we know, such formulations do not exist in the loop expansion (for previous attempts, see \cite{selftuning,degrav}).
As we see by reversing the loop counting argument, the problem is that a theory at a fixed loop order would need to have a very exotic dependence on the Lagrange multipliers in order to restore the simple overall dependence after the resummation of all the loop corrections, which yields the vacuum energy cancellation.
Worse yet, this dependence would have to be theory dependent and especially sensitive to the details of its UV sector, completely contrary to the naturalness criteria. Indeed, changing a theory in the UV, say by adding a very heavy field, would change almost nothing -- except the cosmological constant! -- at low energies, inducing only small corrections to all local observables.
However the cosmological constant changes would be dramatic: the structure of the UV corrections to the vacuum energy would be completely different. 
In other words, two theories can look exactly the same in the IR, but the radiative corrections to their cosmological constants will be wildly different if their spectra differ on the heavy end.
Note that the dependence of the vacuum energy sequester on the Lagrange multipliers looks simple at any given loop order, however the specific form changes as higher order corrections are added. Nevertheless the cancellation of vacuum energy by the dynamics is independent of this since the vacuum energy contributions always depend on the Lagrange multipliers in the exact same way, independent of the loop order.
Thus vacuum energy sequester models and their generalizations provide a simple example of perturbatively reliable algorithms for canceling vacuum energy.

\section{Back to sequester: global, local and hybrid} \label{sec:local}

As we stressed above, the problems with radiative stability that may plague proposals using global constraints to address the cosmological constant problem can be cured by straightforward modifications.
In turn these modifications seem to link them to vacuum energy sequestering, or their generalizations.
A simple illustration of generic features of global vacuum energy sequestering is provided by
\be
S = \int d^4 x \sqrt{ g} \left[ \frac{\kappa_0^2}{2}  R   -\Lambda_0- {\cal L}(g^{\mu\nu} , \Phi) \right] -\Lambda \int d^4 x \sqrt{ g}+\sigma_1 \left(\frac{  \Lambda}{ \mu^4}\right) S_1+\theta S_g[g_{\mu\nu}]+ \sigma_2\left(\theta \right) S_2 \, ,
\ee
where $\Lambda_0$ and $\kappa_0^2$  are fixed parameters, and $\Lambda, \theta$ are two global variables.
The physical cosmological constant and gravitational coupling depend on the specific form of $S_g$ and the interplay between global variables in the system.
We require $S_1$ and $S_2$ to have vanishing variation with respect to the metric or fields contained within the matter Lagrangian, $\frac{\delta S_{1,2}}{\delta g^{\mu\nu}}=\frac{\delta S_{1,2}}{\delta \Phi}=0$. Hence $S_1$ and $S_2$ are either purely topological, or else only having non-vanishing variation with respect to hidden sector fields (that are locally decoupled from the spacetime metric). Furthermore, $S_1$ should be non-vanishing on-shell in order to avoid an unphysical constraint on the spacetime volume. Suitable examples for $S_1$ and $S_2$ include
\begin{itemize}
\item constants
\item$\int F$, where $F=dA$ is a four form field strength
\item $\int \text{tr}(G \wedge G)$, where $G$ is a two form field strength
\item  $\int d^4 y \sqrt{f}\left[ \alpha R(f)+\beta\right]$, where $f_{\mu\nu}$ is a hidden sector metric and $\alpha, \beta$ are dimensionful constants.
\end{itemize}
The form of $S_g$ should not allow for a significant departure from Einstein gravity. It should also be such that the variation with respect to $\theta$ directly constrains the scale dependent part of the geometry. The most conservative choices would be to take $S_g$ to be the Einstein-Hilbert action as in most versions of vacuum energy sequestering \cite{kalpad1,kalpad2,kalpad3,kalpad5,kalpad6}, or the integral of Gauss-Bonnet, as in the most recent version designed to address the problem of graviton loops \cite{kalpad4}.

To illustrate the dynamics of cancelation, we ignore the graviton loops and limit to the case when $S_g=\int d^4 x \sqrt{g}\left[\frac{M^2}{2} R-\alpha M^4\right]$, for some fixed dimensional scale $M$ and a constant $\alpha$. In this case, the metric variation yields the following Einstein's equation
\be
(\kappa_0^2+\theta M^2) G_{\mu\nu}=-(\Lambda_0+\Lambda+\alpha \theta M^4) g_{\mu\nu} +T_{\mu\nu}.
\ee
Note that $S_1$ and $S_2$ do not gravitate and so they do not contribute to this equation. Variation with respect to the global variables yields the following global constraints
\be
\int d^4x \sqrt{g}=\frac{\sigma_1'}{\mu^4}S_1 \, . \qquad \int d^4 x \sqrt{g} \left[\frac{M^2}{2}R- \alpha M^4 \right]=-\sigma_2'S_2 \, .
\ee
The ratio of these two equations yields
\be
\frac{M^2}{2} \langle R \rangle=\alpha M^4-\mu^4 \frac{\sigma_2'S_2}{\sigma_1'S_1} \label{<R>} \, .
\ee
After taking the trace and  spacetime average of Einstein's equation, and substituting in (\ref{<R>}), we obtain the following effective gravity equation
\be
M^2_\text{eff} G_{\mu\nu}=T_{\mu\nu}-\frac14 \langle T \rangle g_{\mu\nu}- \Delta \Lambda g_{\mu\nu} \, ,
\ee
where $M^2_\text{eff}=\kappa_0^2+\theta M^2$ is the observed Planck scale, and
\be
\Delta \Lambda=\frac14 M_\text{eff}^2 \langle R \rangle=\frac12 \alpha M^2 M_\text{eff}^2-\frac12 \mu^4 \frac{M_\text{eff}^2 \sigma_2'S_2}{M^2\sigma_1'S_1 } \, .
\ee
Separating the source into vacuum energy and local excitations, $T_{\mu\nu}=-V_{vac} g_{\mu\nu}+\tau_{\mu\nu}$ we see that the former drops out at each and every order in the loop expansion:
\be \label{Eeqn}
M^2_\text{eff} G_{\mu\nu}=\tau_{\mu\nu}-\frac14 \langle \tau \rangle g_{\mu\nu}-\Delta \Lambda g_{\mu\nu} \, .
\ee

The contribution from $\Delta \Lambda$ can be radiatively stable. Note that the first term is just a number.
It is automatically radiatively stable, and can be arbitrarily small after a suitable choice of $\alpha$.
This would be a tuning, to be sure, but once made loops would never spoil it.
For the second term, the topological terms $S_{1,2}$ are not UV sensitive if they are a pure boundary term or simply a constant.
Next, if $M$ and $\kappa_0$ are Planckian, the measured value of $\theta$ is ${\cal O}(1)$, with radiative corrections being at most ${\cal O}(1)$, as long as the field theory cut-off is near the Planck scale. So as long as $\sigma_2$ is a sufficiently smooth non-linear function, it will be radiatively stable. Similarly, if $\mu$ is near the field theory cut-off, the argument of $\sigma_1$ is radiatively stable, and for a sufficiently smooth non-linear function, so is $\sigma_1$.

While such a simple setup clearly shows how in the end one separates the vacuum energy from the visible sector, it would be interesting to explore the impact of different choices of $S_1$ and $S_2$. 
Note that we have displayed a generic class of theories for which the Standard Model vacuum energy is sequestered at each and every order in loop perturbation theory.
We also emphasize that by choosing $S_g$ to be the integral of the Gauss-Bonnet invariant, we generate a class of models for which vacuum energy loops including gravitons may also be sequestered.
Further, while the residual renormalized cosmological constant, which sets the background curvature of the vacuum geometry, can be small, models which generically predict that the curvature is small would be very interesting.

The global terms $\propto \sigma$ that violate additivity of the action, in apparent\footnote{The conflict might be avoided if these terms arise from non-perturbative effects, in which case the action containing them is not really the phase of the initial path integral, but its saddle point, as noted in the previous section. However since such techniques are not really under control, we do not pursue this possibility here.} conflict with quantum mechanics are perhaps the most unusual
features of global vacuum energy sequester \cite{kalpad1,kalpad2}.
The generalizations we have outlined here suggest ways out of this difficulty, as long as $S_1$ and $S_2$ are integrals over the same spacetime as the visible sectors, as opposed to simple constants as in the original formulation.
Indeed, if $S_1$ and $S_2$ are integrals of 4-form field strengths, $S_i=\frac{1}{4!} \int F^{(i)}_{\mu\nu\alpha\beta}dx^\mu dx^\nu dx^\alpha dx^\beta$, where $F^{(i)}_{\mu\nu\alpha\beta}=4 \partial_{[\mu} A^{(i)}_{\nu\alpha\beta]}$, we can promote the global scalar variables to local degrees of freedom, yielding the manifestly local formulation of vacuum energy sequester  \cite{kalpad3}
\be
S=\int d^4 x \sqrt{ g} \left[ \frac{\kappa^2(x)}{2}  R   -\Lambda(x)- {\cal L}(g^{\mu\nu} , \Phi) \right] +\int \sigma_1 \left(\frac{  \Lambda}{ \mu^4}\right)  F^{(1)}+ \int \sigma_2\left(\frac{\kappa^2}{M^2} \right)  F^{(2)} \, .
\ee
In order to make contact with the notation in \cite{kalpad3}, we have dropped $\kappa_0^2$, $\Lambda_0$,  $\alpha$, and identified $\kappa^2=\theta M^2$. Variation with respect to the 3-forms $A^{(i)}_{\nu\alpha\beta}$ now forces $\kappa^2$ and $\Lambda$ to be constant, and the effective gravitational dynamics can be shown to take the form of  (\ref{Eeqn}) with $M^2_\text{eff}=\kappa^2$, and $\Delta \Lambda =-\frac12 \mu^4 \frac{\kappa^2  \sigma_2'\int F^{(2)}}{M^2\sigma_1' \int F^{(1)} }$. These equations are radiatively stable, becoming completely insensitive to vacuum energy loops in the limit that the flux of $F^{(2)}$ goes to zero.

As noted in \cite{kalpad5}, in the local formulation we impose Dirichlet boundary conditions on the 3-forms and the {\it Einstein} frame metric, and (trivially) Neumann boundary conditions on the scalars in order to retain their global variations. With this at hand, we can integrate out the 4-forms by fixing their flux and retaining only the global variation of the scalars, in effect ``truncating" the theory to its global limit. Alternatively, we can integrate out the scalars, so that the action remains local,
\be
S=\int d^4 x \sqrt{g} \left[ {\cal F}_1 \left(- \frac{\mu^4}{\star F^{(1)}}\right) \star F^{(1)}+{\cal F}_2 \left(  \frac{\frac12 M^2 R}{\star F^{(2)}}\right) \star F^{(2)}- {\cal L}(g^{\mu\nu} , \Phi) \right] \, ,
\ee
where $\star$ corresponds to the Hodge dual operation, and we use the Legendre transform of $\sigma_i( z_i)$ given by ${\cal F}_i (p_i)=z_i p_i-\sigma_i$, where $p_i=\sigma_i'( z_i)$. Clearly such a description does not exist for linear $\sigma_i$, which in any case brings in additional fine tunings \cite{kalpad5}.

How do we recover the global constraints in this formulation, given only the form of ${\cal F}_i(p_i)$? The trick is to use the conjugate variable to $p_i$, which we denote $z_i={\cal F}_i'(p_i)$, and identify the Legendre transform of ${\cal F}_i(p_i)$, which is just $\sigma_i(z_i)$. This implies $p_i=\sigma_i'( z_i)$. Since $p_1=-\frac{\mu^4}{\star F^{(1)}}$ and $p_2= \frac{\frac12 M^2 R}{\star F^{(2)}}$, we obtain the following  equations
\be
F^{(1)}=\frac{1}{\sigma_1'} \mu^4 \sqrt{g} d^4 x, \qquad \qquad F^{(2)}=-\frac{1}{2  \sigma_2'}   M^2 R \sqrt{g} d^4 x \, .
\ee
Integrating these over the cosmic worldvolume effectively yields the global constraints.

We can obtain an interesting hybrid of the local and global formulations of vacuum energy sequester by combining the two procedures described above, integrating out the scalar $\Lambda$ and the 3-form $A^{(2)}_{\nu\alpha\beta}$. If we assume that the flux of $F^{(2)}$ is vanishing, this yields a model similar to our improvement (\ref{crmod}) of the radiatively unstable proposal of \cite{carrollremmen}. The theory is given by
\be\label{TheHybrid}
S=\int d^4 x \sqrt{g} \left[- \mu^4 \epsilon \left( \frac{\star F^{(1)}}{\mu^4} \right) +\frac12 \kappa^2 R- {\cal L}(g^{\mu\nu} , \Phi) \right] \, ,
\ee
where $\epsilon(x)=-x{\cal F}_1(-1/x)$. Although we allow for global variation of the constant $\kappa^2$, this action is additive. The connection to  (\ref{crmod}) is easily obtained by setting $\sigma_1(z)^2=-2z$, from which we obtain ${\cal F}_1(p_1)=1/2p_1$, and so $\epsilon(x)=\frac12 x^2$. From this perspective, it is tempting to identify $\kappa^2$ with a very heavy dilaton field.

Where could such structures come from? Currently string theory provides a powerful framework for UV completions of local QFTs, and we might find hints for how to develop microscopic formulations of vacuum energy sequester
by combining string theory motivated top-down reasoning with bottom-up EFT constructions pursued here.
String theory provides the landscape of quantized fluxes and branes as one mechanism for explaining the observed small positive cosmological constant. It should be taken together with eternal inflation, which gives ways
to scan the landscape and apply the anthropic principle \cite{boupo}. Vacuum energy sequestering is an alternative mechanism within EFT allowing degravitation of the vacuum energy up to some cut-off scale. These two approaches may not contradict each other. It is possible that sequester provides an additional degree of stability 
and predictivity in the string landscape, perhaps linking other sectors to the cosmological constant, similarly to recent ideas of \cite{stanford}. On the other hand, it may also be an avatar of some intrinsically stringy dynamics, as yet unknown, which provide a cosmological constant cancellation without resorting to anthropics.
Anthropics may still be necessary to fix the cosmological initial conditions required to -- for example -- trigger inflation and explain our universe and structures that inhabit it. But the cosmological constant might not be a random boundary variable.

At present we do not know which of these possibilities is true, and so it seems reasonable to explore the string 
arsenal and look for non-quantized 4-forms needed for vacuum energy sequestering,  and their coupling structures either in part (say to the matter sector) or in full including gravity. With this in mind we wish to speculate on how  sequestering might arise in the low-energy approximations to string theory. For this purpose, we observe that type II string theories in the 10D string frame yield a low-energy supergravity effective action which contains structures resembling a 10D version of \eqref{TheHybrid}. To see that, 
one identifies $\kappa=\exp(-\phi)$, imagining the dilaton to be very heavy so that its kinetic terms are negligible, and supplies the various RR-sector field strengths with the structure $\epsilon(x)=\frac12 x^2$.
As their 10D bulk kinetic terms appear in the 10D type II supergravity action without the $\exp(-2\phi)$ prefactor, they nicely fit into the $\epsilon(x)$ piece of \eqref{TheHybrid}. Then ${\cal L}(g^{\mu\nu},\Phi)$ matches the NSNS-sector fields as well as any D-brane sector potentially providing the gauge fields and chiral matter of the SM. Clearly,
the form sectors need to have nontrivial topological properties in order to evade the need to tune the fluxes
generated by the usual local sources.

 If we now envision compactification of the 10D type II actions with these identifications to 4D and using various dualities on the RR-sector p-form field strengths, some of them may provide 4-form field strength pieces for the $\epsilon(x)$-piece in the resulting 4D effective action. To be sure, this argument is really just a caricature at this stage. However it is conceivable that some ingredients in the action \eqref{TheHybrid} might arise from string compactification. As our sequestered theories here, including the above hybrid, are effective theories which in reality should describe a complete UV regulator that cancels the cosmological constant, a
relation to string theory might  provide such a link. Since the cosmological constant in full string theory manifestly vanishes~\cite{Kachru:1998hd,Kachru:1998pg} the string theory dynamics would play the role of the UV regulator above the KK/string scale, whose low energy avatar might be sequestering.

\section{Summary}

The list of proposals for taking on the cosmological constant problem that rely on the imposition of  global constraints is growing.
In this paper we have assessed and compared some examples of such proposals \cite{tsey,kalpad1,kalpad2,westphal,carrollremmen}.
Among these we find that the vacuum energy sequestering \cite{kalpad1,kalpad2} and its relatives are a good starting point for a perturbative formulation of a theory which cancels the vacuum energy contributions at an arbitrary loop level using a dynamical cosmological constant counterterm.
In the examples considered in \cite{tsey,carrollremmen}, radiative corrections are not canceled, and the theory must be retuned order by order in the loop expansion.
If these theories were full effective actions where the corrections are summed up this could be avoided, but then one has to understand where such effective actions come from, a question currently completely open.

While perturbative stability can be restored by relatively simple ``tweaks", such modifications generically convert the theory into a generalized version of vacuum energy sequester.
We have suggested some simple generalizations of global vacuum energy sequester.
Two key ingredients are required: (1) a cosmological constant counterterm allowed to vary globally and (2) a purely geometrical global constraint.
In the Jordan gauge, the latter should  constrain only the Ricci scalar averaged over spacetime.
If, as in \cite{carrollremmen}, one contaminates this constraint by the matter sector, one can run into mismatch at higher loop orders.
In the examples in section \ref{sec:local}, we constrain the spacetime average of the Ricci scalar directly.
This suffices to cancel vacuum energy contributions from matter loops at all orders in perturbation theory.
By imposing a global constraint on a purely {\it topological} invariant,  corresponding to the spacetime average of the Gauss-Bonnet combination, we can extend the cancellation mechanism to take care of loops that also involve virtual gravitons \cite{kalpad4}.

With the use of sourceless 4-form field strengths, one could be able to develop a manifestly local version of vacuum energy sequester, where the global constraints are understood as local field equations \cite{kalpad4,kalpad3}.
In these cases, the spacetime average of the Ricci scalar is constrained by the flux of the 4-form, which is UV insensitive and set by the boundary data.
We have also seen how to develop a curious hybrid of the local and global formulations: here the cosmological constant counterterm corresponds to the dynamics of a 4-form field strength, and the Planck mass acts as a global variable constraining the spacetime average of the Ricci scalar to vanish.
This hybrid is described by an additive action, the cancelation is radiatively stable and the cosmological constant counterterm is picked to exactly cancel the regularized vacuum energy at any order in the loop expansion. Clearly this suggests intriguing possibilities and deserves further attention.

\section*{Acknowledgements}
We would like to thank Asimina Arvanitaki, Wilfried Buchmuller, Ed Copeland, Savas Dimopoulos, Kiel Howe, Renata Kallosh, Matt Kleban, Albion Lawrence, Andrei Linde, Fernando Quevedo, Paul Saffin, Lorenzo Sorbo and John Terning for useful discussions.  NK is supported in part by the DOE Grant DE-SC0009999. AP was funded by an STFC consolidated grant and a Leverhulme Research Project grant. DS acknowledges the Dutch funding agency `Netherlands Organisation for Scientific Research' (NWO) for financial support. The work of A.W. is supported by the ERC Consolidator Grant STRINGFLATION under the HORIZON 2020 contract no. 647995. 
GZ is supported by John Templeton Foundation grant 60253.

\end{document}